\documentclass[showpacs,aps,prl,twocolumn,superscriptaddress,10pt]{revtex4-2}
\usepackage{graphicx} % Include figure files
\usepackage{bm}% bold math
\usepackage{color}

\usepackage{amsmath}
\usepackage{amssymb}
\usepackage{enumerate}
\usepackage{xspace}
\usepackage{mathrsfs}
\usepackage{mathptmx}
\usepackage[colorlinks=true,linkcolor=blue,citecolor=blue,urlcolor=blue]{hyperref}
\usepackage{orcidlink}
\usepackage[utf8]{inputenc}

\newcommand{\ku}{\ensuremath{k_{\mathrm{u}}}\xspace}

\newcommand{\vampire}{\textsc{vampire} }

\begin{document}

\title{Ultrafast imprinting of fused ferromagnetic/antiferromagnetic skyrmions}

\author{Coline~Thevenard\orcidlink{0009-0003-3664-7519}}
\email{coline.thevenard@cea.fr}
\affiliation{Univ. Grenoble Alpes, CNRS, CEA, Grenoble INP, IRIG-Spintec, F-38000 Grenoble, France}
\author{Carenza~E.~Cronshaw\orcidlink{0000-0001-8188-1556}}
\affiliation{School of Physics, Engineering and Technology, University of York, York YO10 5DD, United Kingdom}
\author{Romain Tahay\orcidlink{0009-0003-8013-8017}}
\affiliation{Univ. Grenoble Alpes, CNRS, CEA, Grenoble INP, IRIG-Spintec, F-38000 Grenoble, France}
\author{Daria~Gusakova\orcidlink{0000-0002-5111-7079}}
\affiliation{Univ. Grenoble Alpes, CNRS, CEA, Grenoble INP, IRIG-Spintec, F-38000 Grenoble, France}
\author{Vincent~Baltz\orcidlink{0000-0002-7637-1938}}
\email{vincent.baltz@cea.fr}
\affiliation{Univ. Grenoble Alpes, CNRS, CEA, Grenoble INP, IRIG-Spintec, F-38000 Grenoble, France}
\author{Richard~F.~L.~Evans\orcidlink{0000-0002-2378-8203}}
\email{richard.evans@york.ac.uk}
\affiliation{School of Physics, Engineering and Technology, University of York, York YO10 5DD, United Kingdom}
\begin{abstract}
Antiferromagnetic skyrmions are promising topologically protected entities for robust storage and THz transport of information. Their generation has so far relied on slow techniques or on specific magnetic structures, which limit their integration into spintronic devices. Here, we demonstrate the imprinting of antiferromagnetic skyrmions within a few picoseconds in a coupled ferromagnetic/antiferromagnetic bilayer with a thermal laser pulse. We identify a threshold energy required to initiate the fused skyrmion phase and show that this can be modulated by a magnetic field, enabling deterministic nucleation of antiferromagnetic skyrmions. Compared to conventional slow field-cooling processes to control antiferromagnets we show that laser-induced skyrmion imprinting can be orders of magnitude faster and provides an experimentally feasible route towards ultrafast creation and manipulation of antiferromagnetic skyrmions.
\end{abstract}

%\pacs{75.10.Hk,75.20.-g,75.50.Ss,75.60.Jk,75.78.Jp}
\maketitle

%\section{Introduction}

The intrinsic topological properties of antiferromagnetic (AFM) skyrmions (Sks) confer on them great potential for protected information storage and ultrafast logic computing~\cite{Xia2017,Barker2016,Zhang2016,Shen2018,Potkina2020,Khan2021,Aldarawsheh2024,Lau2025,Berg2026,LeiviskaAS2026}. In particular the strong antiparallel exchange coupling within AFM Sks ensures both ultrafast dynamics in the THz range and remarkable immunity to stray fields. Their magnetic compensation is accompanied by a zero topological charge, giving rise to unique transport properties through the mitigation of the Sk Hall effect~\cite{Barker2016,Zhang2016,Bhukta2026} while preserving the topological spin Hall effect \cite{BuhlPSS2017}. Moreover, vanishing net magnetization limits dipolar interactions between Sks potentially enabling higher density compared to their ferromagnetic (FM) counterparts. Despite these intrinsic advantages, the absence of a net magnetization also hinders the control of the AFM order parameter for the nucleation of these non-trivial topological states. Electrical~\cite{Khan2021} and magnetic field~\cite{Berg2026} approaches for nucleation, detection and control of AFM Sks have been demonstrated, but they typically require long timescales and offer limited spatial selectivity. One other possible solution is to couple the AFM to an adjacent FM through interfacial exchange coupling, thereby exploiting the magnetic response of the ferromagnet for nucleation and detection, while retaining the inherent advantages of the AFM Sk. In this case, AFM Sks can be created by thermomagnetic imprinting ~\cite{ThevenardPRB2025,LeiviskaPRB2023,RanaAPL2021}. Despite its broad material applicability, it does not allow for deterministic spatio-temporal control of nucleation suitable for spintronic devices due to the non-locality of the applied magnetic field and the multiple nanosecond cooling time.

An alternative strategy exploits femtosecond laser pulses to efficiently nucleate topological structures~\cite{ZhangMF2023}. The underlying microscopic mechanisms depend on the materials under consideration. In some materials the spins can couple directly to an effective magnetic field via the inverse Faraday effect, as demonstrated in collinear FMs \cite{FlovikPRB2019} and AFMs \cite{KhoshlahniPRB2019}, or via the spatio-temporally singular electrodipolar pulse in a two-dimensional FM~\cite{PolyakovJAP2020}. In other materials, laser-induced Sk formation relies on the coupling of spins to itinerant electrons mediated by thermal excitation~\cite{VinasNPJ2022}. The viability of this approach has been demonstrated both theoretically and experimentally in FM Co and CoFeB~\cite{KernAM2025,OllerosRodriguezN2022,ButtnerNM2021,GerlingerAPL2021,JENL2018}, ferrimagnetic TbFeCo~\cite{Savoini2014,KoshibaeNC2014,FinazziPRL2013}, as well as in AFMs consisting of antiferromagnetically coupled FM planes, such as A-type FeGe \cite{BerrutoPRL2018} and the synthetic AFM [Pt/Co/Ru]$_6$ \cite{JugeNC2022}. 
%Their numerically combined action has finally converted a Sk from the FM to the AFM state of a CrInSe$_3$ monolayer \cite{DouS2025}. 
% Fig. 1: defined here (between intro paragraphs 2 and 3) with [b] so that it
% lands at the foot of the right-hand column of page 1.
\begin{figure}[b]
\centering
\includegraphics[width=0.85\columnwidth]{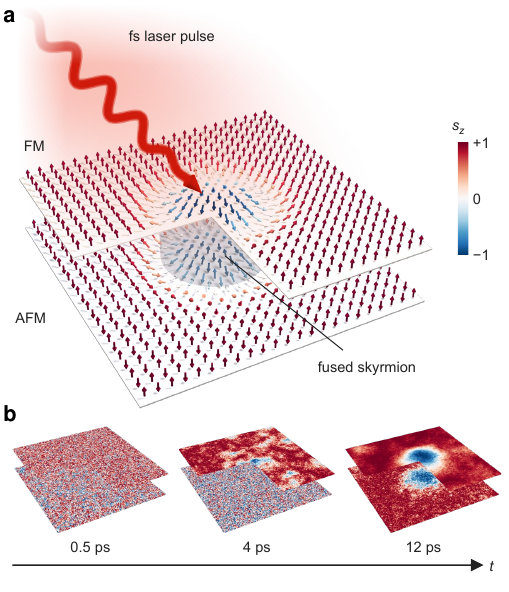}
\caption{Ultrafast imprinting of a fused FM/AFM Sk. (a) A femtosecond laser pulse heats the exchange-coupled FM/AFM bilayer (schematic, not to scale, with one quadrant of the FM cut away). On cooling the FM forms a N\'eel Sk, which the interfacial exchange imprints into the AFM. The colour scale denotes $s_z$ in the FM and the $z$ component of the N\'eel vector in the AFM. (b) Simulated $40 \times 40$ nm$^2$ FM and AFM monolayers (MLs) after a 100-fs, 100 mJ/cm$^2$ pulse, with the AFM replicating the FM Sk by 12 ps.}
\label{fig:nuc-process}
\end{figure}

Laser-induced heating further allows nanoscale control using a near-field transducer together with high energy efficiency due to the low amount of energy per pulse. Crucially, the femtosecond timescale of the excitation is also compatible with the intrinsically ultrafast dynamics of AFMs.

In this letter we employ atomistic simulations to demonstrate the use of ultrafast laser pulses for imprinting Sks from a FM into an adjacent AFM, thereby generating a fused Sk (Fig. \ref{fig:nuc-process}). We identify the optimal conditions for Sk imprint by exploring the interplay between the intrinsic parameters of the magnetic structure and the external parameters of the thermal pulse and an applied magnetic field, and their impact on the topology and stability of the resulting AFM Sks.

% Fig. 2: defined early with [t] so that it floats to the top of page 2.
\begin{figure}[t]
\includegraphics[width=0.45\textwidth, trim=0 0 0 0]{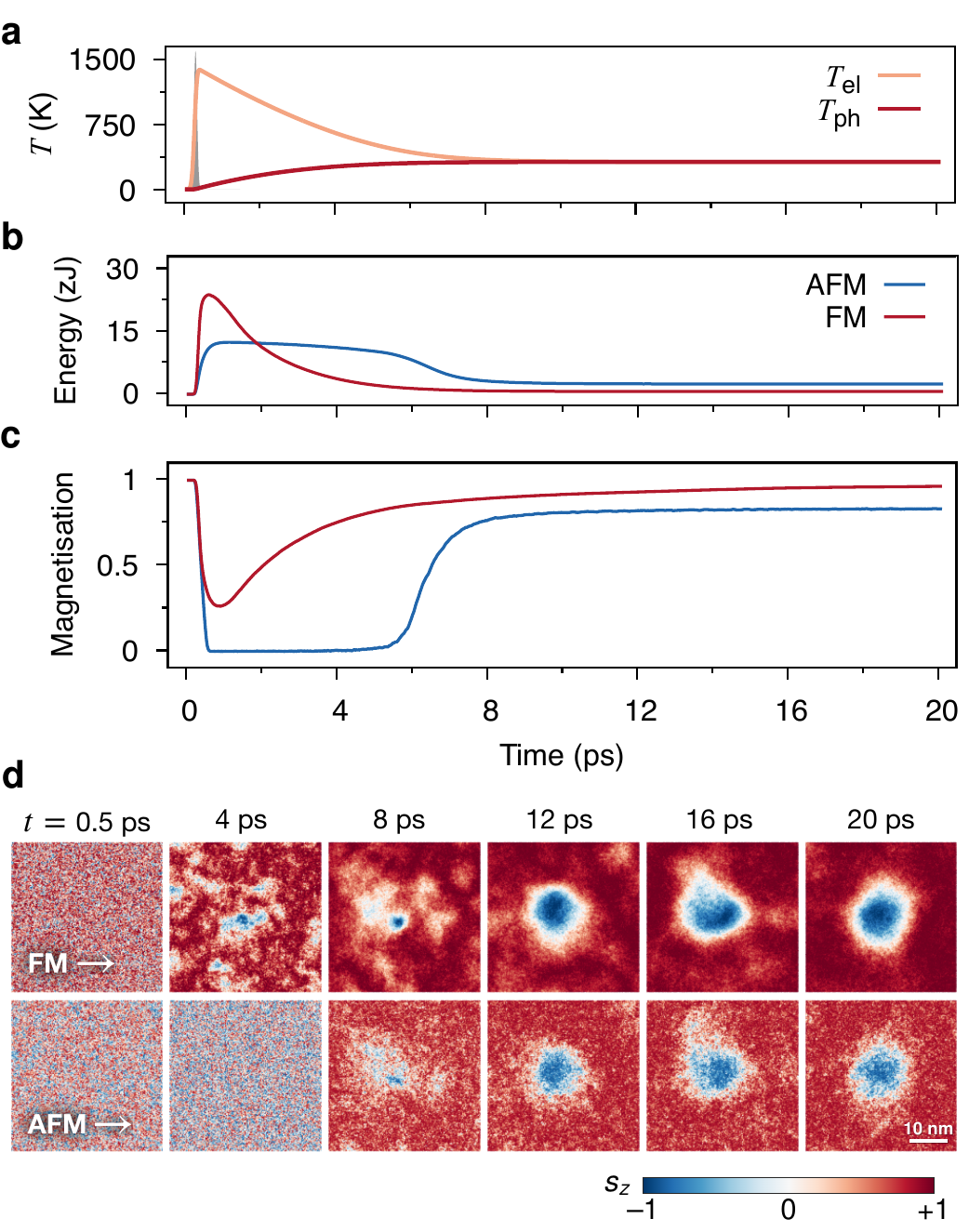}
% TODO: unify temperature notation (T_el/T_ph in the figure, T_e/T_p in the SM).
% TODO: monolayer positions of the snapshots (XX ML / 2ND ML from the FM/AFM interface) in (d).
% A version with a topological-charge panel is parked in Fig2/compose_fig2_with_Q.tex (Fig2/Fig2_with_Q.pdf).
\caption{Temporal evolution of the spin system following a 100-fs, 100 mJ/cm$^2$ thermal excitation. (a) Electron ($T_{\mathrm{el}}$) and phonon ($T_{\mathrm{ph}}$) temperatures, with the temporal profile of the laser pulse shown in grey. (b) Change in energy per spin relative to $t=0$ for the AFM and the FM. (c) Normalized length of the magnetization in the FM and of the N\'eel vector in the AFM. (d) Top-view cross-section snapshots of the FM (top row, 2nd ML from the FM/AFM interface) and of the AFM N\'eel vector (bottom row, 4th ML from the FM/AFM interface). Each snapshot shows a $40 \times 40$ nm$^2$ region of a single ML, the scale bar is 10 nm and the colour scale denotes $s_z$ (the $z$ component of the N\'eel vector in the AFM). The FM Sk forms first and the AFM replicates its topology once N\'eel order is restored.}
\label{fig:syst-evol}
\end{figure}

%\section{Methodology}
The numerical study models $N=2.1 \times 10^{6}$ spins in a rocksalt crystal structure with lattice constant of 4.17~$\AA$, covering a volume of $100 \times 100 \times 1.8765$~nm$^{3}$, unless otherwise specified. The layered system is an AFM (8.34)/NM (2.085)/FM (6.255)/HM (2.085 $\AA$) stack with a NM ML mixed with 25 $\%$ of FM atoms for FM/AFM exchange coupling, as used previously~\cite{ThevenardPRB2025}. The atomistic simulations were conducted using the \vampire software package \cite{EvansAtomisticPRB2015}. The spin Hamiltonian
\begin{equation}
\begin{split}
\mathscr{H} = &
-\sum_{i<j} J_{ij} \textbf{S}_i \cdot \textbf{S}_j -\sum_{i<j} \textbf{D}_{ij}\cdot(\textbf{S}_i \times \textbf{S}_j) \\
&
-\ku \sum_i \textbf{S}_{z,i}^{2}
-k_N \sum_{i<j} (\textbf{S}_i \cdot \textbf{e}_{ij})^2 -H_{\mathrm{ext}} \sum_i \mu_{s,i} \textbf{S}_{z,i}.
\end{split}
\label{eq:hamiltonian}
\end{equation}
describes the energy of the spins, where $\textbf{S}_i$ is a unit vector along the spin direction at site $i$ with magnetic moment $\mu_{s,i}$. The first term is the Heisenberg exchange between sites $i$ and $j$, with the exchange constants $J_{ij}$ given in Table~S1 of the Supplemental Material (SM)~\cite{supp}. The exchange is stronger in the FM than in the AFM, and a partial population of NM atoms in the interfacial layer frustrates the exchange at the FM/AFM interface to make the imprinting efficient. The second term is the interfacial Dzyaloshinskii-Moriya interaction (DMI) at the HM/FM interface, with DMI vector $\textbf{D}_{ij}$, which stabilizes the Sks. The third and fourth terms are the uniaxial anisotropy with constant $k_u$ and the N\'eel surface anisotropy with constant $k_N$, where $\textbf{e}_{ij}$ is the unit vector joining neighboring sites $i$ and $j$. The final term is the Zeeman interaction with an external magnetic field $H_\mathrm{ext}$, which is zero unless otherwise specified. The structural and magnetic parameters are adopted from those optimized for  AFM Sk imprinting using a field-cooling procedure \cite {ThevenardPRB2025}. To enhance the effective interfacial coupling the thickness of the AFM layer is reduced to 4 nm, and the anisotropy increased to help stabilize the imprinted Sks ~\cite{JohnsonRPP1996}. The temporal response of the system to a laser pulse is determined by solving the stochastic Landau-Lifshitz-Gilbert equation through the Heun predictor-corrector method~\cite{EvansAtomisticPRB2015}. We use characteristic damping coefficients $\lambda_{FM}=0.05$~\cite{FujitaJMMM2008} and $\lambda_{AFM}=0.01$~\cite{SimensenPRB2020}. The effective field incorporates the effects of magnetic interactions given by the derivative of the spin Hamiltonian and the thermal bath using the Langevin dynamics approach detailed in the SM~\cite{supp}. The effects of the laser pulse are simulated using a two-temperature model with a Gaussian time profile (SM). Quantum effects are incorporated phenomenologically with spin-temperature rescaling~\cite{EvansQuantitativePRB2015}. All simulations were performed on the UK national supercomputer \textsc{archer2}~\cite{archer2} and on the Luxembourg national supercomputer \textsc{meluxina}, using 8 nodes (1024 cores) on each.

%\section{Results}

%\subsection{Ultrafast thermal imprinting of fused Sk}

The nucleation of fused Sk by an ultrafast thermal pulse is depicted in Figs.~\ref{fig:nuc-process} and~\ref{fig:syst-evol}. The laser excitation leads to a rapid increase of the electron temperature which is coupled to the spin system, as shown in Fig.~\ref{fig:syst-evol}(a). The excitation also modifies the magnetic energy landscape (Fig. \ref{fig:syst-evol}(b)) and disrupts the pre-established FM and AFM order, characterized by the respective normalized magnetization $\textbf{m}=\frac{1}{N_{FM}}\sum_{i\in \text{FM}} \textbf{S}_i$ and N\'eel vector $\textbf{l}=\frac{1}{N_{AFM}}\left( \sum_{i \in AFMb}\textbf{S}_i - \sum_{i \in AFMa}\textbf{S}_i \right)$ where a and b are the AFM sublattices (Fig. \ref{fig:syst-evol}(c)). This reconfiguration drives the formation of non-trivial spin textures quantified by the topological charge $Q$, sum of the topological charge density $q=\frac{1}{4\pi}\sum_{(i,j,k)\in\{(1,2,3),(1,3,4)\}}{2\text{tan}^{-1}\left(\frac{\textbf{S}_i\cdot(\textbf{S}_j\times\textbf{S}_k)}{1+\textbf{S}_i\cdot\textbf{S}_j+\textbf{S}_i\cdot\textbf{S}_k+\textbf{S}_j\cdot\textbf{S}_k}\right)}$ per square cell delimited by sites $i \in [\![1,4 ]\!],$ from the Berg and Lüscher's approach~\cite{BergLuscher1981}. We evaluate $Q_{\mathrm{FM}}$ for the FM and $Q_{\mathrm{AFM}}$ for the AFM sublattice b, whose spins follow the N\'eel vector, since the two AFM sublattices carry charges of opposite sign.

The observed spin dynamics can be broken down into three successive stages. First, we start with a saturated system at $t = 0$ to render a skyrmionic metastable state thermally accessible from the domain wall ground state. Imprinting assisted by a strong magnetic field or a transfer of AFM spin torque from a reservoir devoid of DMI can align the magnetization and the Néel vector out-of-plane such that $\textbf{m},\textbf{l}=\textbf{\^{z}}$ at 0 K. The uniformity of the spin orientation translates into a zero topological charge. We then apply an ultrashort heat pulse representing the laser which brings about transient electronic heating. The spin response to the stochastic field is delayed by around 
%$\underset{\tau}{\text{argmax}} \int_{0}^{20 \text{ps}}T(t) E_{deposited}(t+\tau)dt=65$ 
65 fs before significantly competing with exchange and anisotropy interactions. Thermal excitation drives the system beyond a blocking temperature $T_{B1}=1170$ K. $T_{B1}$ is above the Néel temperature of the AFM $T_N=505$ K, causing it to transition to a paramagnetic state where $\textbf{l}\sim \textbf{0}$, but below the Curie temperature of the FM $T_C=1339$ K, preserving the FM component $\textbf{m}=0.26~\textbf{\^{z}}$ without a phase collapse. Thermal fluctuations there provoke extraordinary angular variations between neighboring spins, which prevent an exact instantaneous topological definition. Finally the electrons transfer their energy to the phonon bath. The FM then reorders its magnetization $\textbf{m}=(0.03,0.02,0.94)$ into a Sk with a topological charge of 1. Around 6 ps later upon cooling below $T_N$, the AFM restores the N\'eel order with $\textbf{l}=(0.02,0.02,0.82)$ along $\textbf{m}$, and further replicates the FM topology at $T = 325.75$~K. A minor deviation in $m_z$ and $l_z$ persists due to the different magnetic interaction strengths of the two materials. The electron-phonon thermalization is reached at a common temperature of 317 K, the injected energy remaining confined to the system in the absence of extrinsic dissipation. Residual thermal fluctuations, despite their amplitude, do not destabilize the Sk and give rise to no additional nucleation or annihilation events. They can however interfere with the assessment of the topological charge. This dynamic process demonstrates the ability to nucleate strongly coupled fused FM/AFM Sks within a few picoseconds using a laser pulse and significantly faster than using conventional field-cooling or current-driven spin-orbit torque (SOT), in which a charge current in an adjacent HM exerts a torque on the magnetization~\cite{Manchon2019}.

\begin{figure}[!tb]
\includegraphics[width=\columnwidth]{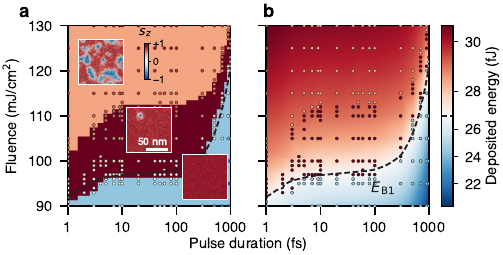}
\caption{Diagram of the hybrid (coral), Sk (brown) and saturated (blue) phases (a), and deposited energy (b) as a function of thermal pulse fluence and duration. Simulation data points are colored by their energy (a) and phase (b). The dashed black curve, labelled $E_{B1}$, delineates the topological energy barrier for fused Sk nucleation. Insert: Top view cross-section snapshots of the spin textures in the AFM (4th ML from the FM/AFM interface), with the colour scale denoting $s_z$.}
\label{fig:phase-diagram}
\end{figure}

\begin{figure}[ht]
\includegraphics[width=\columnwidth]{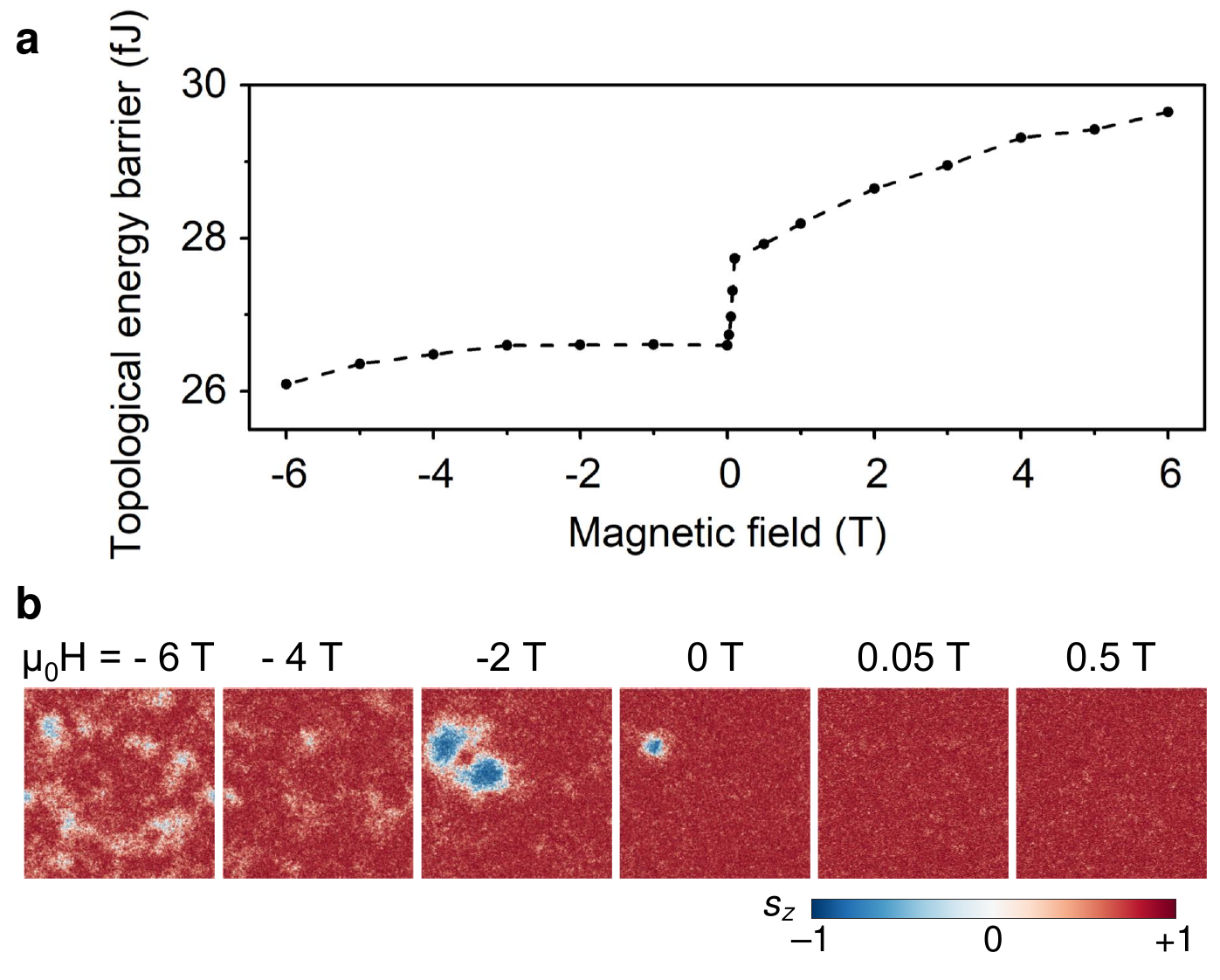}
% TODO (Coline): pulse duration used for Fig. 4.
\caption{(a) Magnetic field dependence of the topological energy barrier $E_{B1}$ for a $200 \times 200$ nm$^2$ film normalized to a $100 \times 100$ nm$^2$ area, defined as the energy difference between the initial saturated state and the transition state at the boundary between the saturated and Sk phases. Negative fields are antiparallel to the initial magnetization. (b) Top view cross-section snapshots of the spin textures in the AFM (4th ML from the FM/AFM interface), with the colour scale denoting $s_z$.}
\label{fig:H-barrier}
\end{figure}

%\subsection{Thermal pulse optimisation}
To better understand the parameter space for successful nucleation we performed a series of calculations for different laser power and pulse duration to construct a phase diagram of the final magnetic state of the system shown in Fig.~\ref{fig:phase-diagram}. To reduce finite-size effects and stochasticity in the simulated spin dynamics the simulated area is increased to $200 \times 200$ nm$^2$. All deposited energies and energy barriers in Figs.~\ref{fig:phase-diagram} and~\ref{fig:H-barrier}(a) are then normalized to a $100 \times 100$ nm$^2$ area. We classify the final state of each simulation from the topological charges of the FM and AFM and the spatial distribution of the topological charge density. The saturated phase has $Q_{\mathrm{FM}}=Q_{\mathrm{AFM}}=0$ and a uniform order parameter along $+\hat{\textbf{z}}$. 
In the fused Sk phase, $q$ forms a disk centred on each Sk peaked at its core \cite{ThevenardPRB2025}. Its integral over each texture equals unity, so that $Q_{\mathrm{FM}}=Q_{\mathrm{AFM}}\in\mathbb{N}^{*}$ directly counts the Sks. The FM and AFM textures share the same vorticity set by interfacial DMI and the same polarity inherited from the initial order established under a field that is switched off for nucleation. It is this initial order that imposes $Q_{FM,AFM}>0$. Reversing the sign of $Q_{FM,AFM}$ would require erasing this residual effect of the field which instead prevents Sk formation altogether. We assign all remaining states to a hybrid phase in which $q$ exhibits a broad distribution of both signs: Sks coexist with elongated domains, skyrmoniums and textures of higher-order charge. In this regime, $Q_{FM,AFM}$ may be non-integer values, reflecting textures with incomplete winding like isolated merons or domains. For high fluences and long pulses the final temperature is too high for the charge to be evaluated. We therefore classify the textures after an additional cooling step until the charge has stabilized.

From the phase diagram of the adjustable variables of the thermal pulse a unique window for the fused Sk nucleation emerges. For intermediate pulse durations the Sk phase is bounded by the isoenergies $E_{B1}=27$ fJ and $E_{B2}=29$ fJ. We define the topological energy barrier $E_{B1}$ as the energy difference between the initial saturated state and the transition state at the boundary between the saturated and Sk phases, shown by the black curve in Fig.~\ref{fig:phase-diagram}. This definition holds where the final temperature is low enough for the charges, and so the phase boundary, to be calculated. 
%\textcolor{red}{The extracted estimate remains above the theoretical barrier $E_{B1,th}=???$ due to these effects and the thermal annihilation in play for long pulses \cite{OllerosRodriguezN2022}}.

For $E_{deposited}<E_{B1}$, the pulse energy is insufficient to destabilize the pre-established uniform state. Long pulses further attenuate the excited spin dynamics by electron-phonon mediated energy dissipation. This regime may nevertheless be exploited to imprint spin textures across weakly coupled magnetic layers. Applied to a FM Sk and saturated AFM, the pulse could raise the AFM energy slightly above $E_{N}=N_{AFM}k_BT_{N}$ with $N_{AFM} = 9 \times 10^5$ the number of AFM atoms, and let it relax toward the Sk texture imposed by the unaltered FM configuration. For $E_{B1}<E_{deposited}<E_{B2}$, fused Sks are nucleated by thermal activated crossing of the topological barrier, with a probability that increases with increasing laser fluence. The phase diagram shows a wider Sk nucleation window for short pulses, which can be explained by the efficient trapping of metastable textures during fast cooling. The nucleation time of the fused Sk lies between 8.4 and 11.4 ps for pulses up to 100 fs and increases to 13.8 ps for a 300 fs pulse (SM Fig.~S1), which suggests that the cooling of the spin system sets the imprinting time for short pulses. Conversely, longer pulses accentuate the thermal annihilation of nucleated spin textures, so that only sufficiently large Sks are stabilized. For larger fluences $E_{deposited} > E_{B2}$, the system transits to a hybrid phase in which Sk bubbles of opposite polarities and domain walls coexist. % Removed (Coline): "Near and beyond E_C = k_B T_C, the system relaxes to its ground state of domain walls in the absence of an external field." This duplicated the field discussion below.
In general we find the nucleation of fused Sks is possible over a wide range of laser fluences and pulse durations, highlighting the robustness of the process and its potential for experimental realisation and applications.

%\subsection{Magnetic field modulation}
While laser pulses can be used to nucleate fused Sks, it is also interesting to explore the possibility of control of the nucleation process through external means such as applied magnetic fields, the inverse faraday effect or an SOT applied as an electrical pulse. For simplicity we consider only a static magnetic field. The effect of an applied magnetic field on the topological energy barrier $E_{B1}$ is shown in Fig.~\ref{fig:H-barrier}. Although the domain wall phase is the ground one at 0 T, the skyrmionic phase becomes energetically favorable within an intermediate field range, broadening the fused Sk nucleation windows, particularly for high fluences or pulses. For strong fields, the saturated state dominates, gradually closing this window. The field orientation strongly modulates this dynamic. If it is antiparallel to the initial order ($H<0$ T), it tends to reverse the spins and thus stabilize the skyrmionic phase. However, its contribution remains marginal at low fields, and becomes significant only at large fields where it effectively competes with the exchange and anisotropy interactions. If the field is parallel ($H>0$ T), the pre-established saturated order is stabilized by compensating for thermal field perturbations, thereby inhibiting the nucleation of fused Sks even under weak fields. At $H = 0$ T, the system transitions between these two states.

%\section{Conclusions}
In conclusion, we have found that ultrafast laser pulses are able to nucleate fused ferromagnetic/antiferromagnetic skyrmions on the picosecond timescale, orders of magnitude faster than existing field-cooling or spin-orbit torque methods. This provides an enticing means of generating antiferromagnetic skyrmions for data storage in antiferromagnetic racetrack memory~\cite{Parkin2008,Duine2018} or neuromorphic computing with control using an external applied field to reset the system to a uniform state or one with fused skyrmions. While the possibility of generating fused skyrmions has been demonstrated theoretically there are many unanswered questions. Ideal materials systems for this approach would be Mn$_2$Au~\cite{Bommanaboyena2021,Jenkins2024} or CuMnAs~\cite{Wadley2016} coupled with a ferromagnetic capping layer of Co/Pt to induce a strong DMI, and also naturally study the dynamics of these coupled skyrmions. Different material systems can also behave in complex ways, particularly considering antiferromagnetic oxides such as NiO and orthoferrites~\cite{Kimel2004}, or non-collinear antiferromagnets such as Mn$_3$Sn~\cite{Ikhlas2017,Tsai2020,Takeuchi2021} and Mn$_3$Ir~\cite{Jenkins2021}. Furthermore, the approach here also allows the creation of purely antiferromagnetic skyrmions, though measurement presents a significant challenge for pure antiferromagnets~\cite{Saidl2017} compared to the fused skyrmions investigated here. 

\section*{Author contributions}
CT and RT performed the atomistic simulations and analysed the data. CT, CEC, DG, VB and RFLE developed the model of the coupled FM/AFM system. CT and RFLE produced the final figures for publication, following their conceptualization and interpretation with DG and VB. VB, DG and RFLE conceived and supervised the project. All authors discussed the results and contributed to the final manuscript.

\section{Acknowledgments}
%\begin{acknowledgments}
The authors would like to thank Alexey Kimel for helpful discussions and Hannah Martin for developing the visualisation of the coupled skyrmions. We would like to acknowledge the provision of computational facilities for this project from the University of York's \textsc{viking} cluster, the UK National Supercomputer \textsc{archer} and Euro HPC Grant EHPC-REG-2025R02-377 and the LuxProvide support teams. The authors gratefully acknowledge the EPSRC Spintronics Network+ Grant No.~EP/V007211/1 and the France 2030 program PEPR SPIN Grants No. ANR-22-EXSP-0003 and ANR-24-EXSP-0002, which supported this work.
%\end{acknowledgments}

\appendix

%\clearpage
%\bibliography{/Users/rfle500/Documents/Work/Papers/Bibliography/library}
%\bibliography{library,local}
\bibliographystyle{apsrev4-2}
\bibliography{library}
%---------------------------------------------------------------------------%

%\begin{thebibliography}{11}%
%\end{thebibliography}%

%\begin{thebibliography}{11}%
%\end{thebibliography}%

\end{document}